\documentclass{aa}
\usepackage{graphicx}

\def\0{\hspace*{0.5em}}
\newcommand{\Rstar}{\ensuremath{\mathrm{R}_{\star}}}
\newcommand{\Rout}{\ensuremath{\mathrm{R}_{\rm out}}}
\newcommand{\Rsun}{\ensuremath{\mathrm{R}_{\odot}}}
\newcommand{\uchii}{UCH{\sc ii}}
\newcommand{\hii}{H\,{\sc ii}}
\newcommand{\bra}{Br$\alpha$}

\newcommand{\brg}{Br$\gamma$}

\newcommand{\pfg}{Pf$\gamma$}

\newcommand{\hu}{Hu(14-6)}

\newcommand{\hei}{\ion{He}{i}}
\newcommand{\hi}{\ion{H}{i}}

\newcommand{\feii}{\ion{Fe}{ii}}
\newcommand{\mgi}{\ion{Mg}{i}}
\newcommand{\mgii}{\ion{Mg}{ii}}

\newcommand{\mum}{\ensuremath{\mu\mathrm{m}}}
\newcommand{\msunyr}{M$_{\odot}$\,yr$^{-1}$}
\newcommand{\Msun}{M$_{\odot}$}
\newcommand{\micron}{\ensuremath{\mu {\rm m}}}
\newcommand{\lam}{$\lambda$}
\newcommand{\vlsr}{\ensuremath{v_{\rm lsr}}}
\newcommand{\kmsec}{\mbox{km\,s$^{-1}$}}

\newcommand{\Lsun}{\ensuremath{\mathrm{L}_{\odot}}}  
\newcommand{\mdot}{\ensuremath{\dot{M}}}              
\newcommand{\rstar}{\ensuremath{\mathrm{R}_{\star}}}  
\newcommand{\teff}{\ensuremath{\mathrm{T}_{\rm eff}}}

\newcommand{\Fnu}{\ensuremath{\mathcal{F}_{\nu}}}       
\newcommand{\rdisk}{\ensuremath{\mathrm{R}_{\rm disk}}}    
\newcommand{\tdisk}{\ensuremath{\mathrm{T}_{\rm d}}}    
\newcommand{\rhozero}{\ensuremath{\mathrm{\rho_{\circ}}}}   
\newcommand{\AV}{\ensuremath{\mathrm{A_{\rm V}}}}       

\begin{document}
\title{The peculiar circumstellar environment of NGC2024-IRS2}
\titlerunning{The peculiar circumstellar environment of NGC2024-IRS2}
\author{A. Lenorzer\inst{1}, 
        A. Bik\inst{1}, 
        A. de Koter\inst{1}, 
        S.E. Kurtz\inst{2},
        L.B.F.M. Waters\inst{1,3}, 
        L. Kaper\inst{1}, 
        C.E. Jones\inst{4}, 
        T.R. Geballe\inst{5}
       }
\authorrunning{A. Lenorzer et al.}
\offprints{A. Lenorzer (lenorzer@astro.uva.nl)} 
\institute{Astronomical Institute "Anton Pannekoek", 
           Kruislaan 403, 1098 SJ Amsterdam, the Netherlands 
      \and Instituto de Astronom\'{i}a, 
           Universidad Nacional Aut\'{o}noma de M\'{e}xico,
           Apdo. Postal 3-72, 58090 Morelia, Mich. M\'{e}xico
      \and Instituut voor Sterrenkunde, K.U. Leuven, 
           Celestijnenlaan 200B, 3001 Heverlee, Belgium 
      \and Department of Physics and Astronomy, 
           The University of Western Ontario, 
           London, Ontario, Canada N6A 3K7
      \and Gemini Observatory, 
           670 N. A'ohoku Place, Hilo, HI 96720, USA
          }
\date{Received; accepted}

\abstract{
We re-examine the nature of NGC2024-IRS2 in light of the recent
discovery of the late O-type star, IRS2b, located 5\,\arcsec\ from
IRS2.  Using L-band spectroscopy, we set a lower limit of \AV\ = 27.0
mag on the visual extinction towards IRS2. Arguments based on the
nature of the circumstellar material, favor an \AV\ of 31.5 mag.  IRS2
is associated with the \uchii\ region G206.543-16.347 and the infrared
source IRAS\,05393-0156. We show that much of the mid-infrared
emission towards IRS2, as well as the far infrared emission peaking at
$\sim$ 100 \mum, do not originate in the direct surroundings of IRS2,
but instead from an extended molecular cloud. Using new K-, L- and
L'-band spectroscopy and a comprehensive set of infrared and radio
continuum measurements from the literature, we apply diagnostics based
on the radio slope, the strength of the infrared hydrogen
recombination lines, and the presence of CO band-heads to constrain
the nature and spatial distribution of the circumstellar material of
IRS2.  Using simple gaseous and/or dust models of prescribed geometry,
we find strong indications that the infrared flux originating in the
circumstellar material of IRS2 is dominated by emission from a dense
gaseous disk with a radius of about 0.6 AU.  At radio wavelengths the
flux density distribution is best described by a stellar wind
recombining at a radius of about 100 AU.  Although NGC2024/IRS2 shares
many similarities with BN-like objects, we do not find evidence for
the presence of a dust shell surrounding this object.  Therefore, IRS2
is likely more evolved.
}

\maketitle

\section{Introduction}

NGC\,2024 is one of the four major nebulae in the nearby giant star
forming complex Orion B. Located at about 363 pc (Brown et al. 1994)
this nearby \hii\ region is particularly active in star formation and
has therefore been extensively studied at many wavelengths. The
optical image of NGC\,2024 (the Flame Nebula) shows a bright
nebulosity with a central elongated obscuration in the north-south
direction.  The heavy extinction renders most of the nebula
unobservable, from the UV to about one micron.

The ionizing source of NGC\,2024 was unknown until very recently (Bik
et al. 2003), although numerous secondary indicators suggested that
the dominant source must be a late-O main-sequence star (e.g. Cesarsky
1977). The ionizing star was searched for in the near-infrared where
it was expected to be the brightest source in NGC\,2024.  This led to
the discovery of IRS2 by Grasdalen (1974). IRS2 has been extensively
studied in the near-infrared (Jiang, Perrier \& L\'ena 1984; Black \&
Willner 1984; Chalabaev \& L\'ena 1986; Barnes et al.  1989; Nisini et
al. 1994), although it was eventually recognized that it is incapable
of powering the \hii\ region (e.g., Nisini et al. 1994). NGC\,2024 hosts
the strong IRAS source 05393-0156 whose colors fulfill the criteria of
Ultra-Compact \hii\ (\uchii) regions as defined by Wood \& Churchwell
(1989).  Kurtz et al. (1994) and Walsh et al. (1998) identified the
compact radio source G206.543-16.347, 72.8\,\arcsec\ from the IRAS
source, but coincident with IRS2.

The dominant ionizing source of NGC\,2024 has recently been identified
as IRS2b. IRS2 is only 5\arcsec\ away from IRS2b, for which an
extinction has been reliably determined (Bik et al. 2003). We address
the nature of the circumstellar material of IRS2 in light of this
discovery. If IRS2 is not the dominant source of ionization of
NGC\,2024, what makes it so bright at infrared wavelengths? Is the
star surrounded by a dust shell and/or a circumstellar disk? Is IRS2 a
massive young stellar object, and if so, do we observe remnants from
the star formation process?

In this study, we make use of spectroscopic observations obtained with
The {\em Very Large Telescope} (VLT) and the {\em United Kingdom
Infrared Telescope} (UKIRT) and photometric measurements from the
literature.  The paper is organized as follows: In
Sect.~\ref{sect_obs} we describe the new spectroscopic observations.
We present the flux density distribution in Sect.~\ref{sect_sed} and
propose a set of diagnostics to investigate the circumstellar material
of IRS2 in Sect.~\ref{sec_tool}. We employ and discuss the merits of
three simple models to fit the observations in Sect.~\ref{sec_discussion}.
We summarize our findings in Sect.~\ref{sec_conclusion}.

\section{Observations}
\label{sect_obs}

\subsection{Observations and data reduction}
\label{sect_red}

    \begin{figure}[t]
    \begin{center}
      \resizebox{8.5cm}{!}{\includegraphics{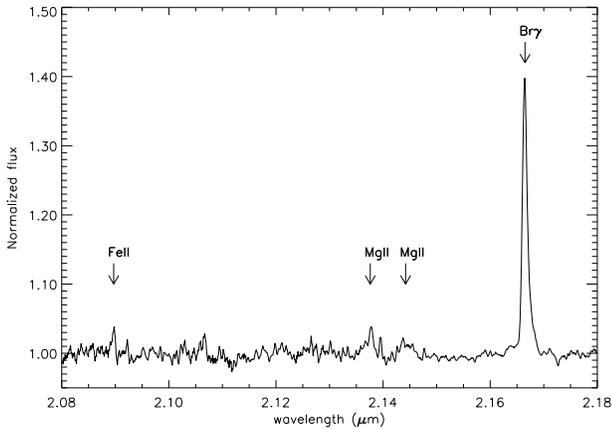}}
      \caption{2.08-2.18 \mum\ spectrum of NGC\,2024 IRS2 obtained
               with ISAAC/VLT
               at $R \simeq 8\,000$. Line identifications 
               and properties are given in Table~\ref{tab_lines}.}
      \label{fig_K}
   \end{center}
   \end{figure}

A 2.08-2.18 \mum\ spectrum of IRS2, with a spectral resolving power $R
\simeq 8\,000$ and signal-to-noise ratio S/N $\sim$ 130, was obtained
using the Infrared Spectrometer And Array Camera (ISAAC) at the VLT
on March 20, 2000.  The spectrum was reduced
using standard procedures.  Telluric absorption lines were removed
using the A2\,V star HD\,39908 observed under identical sky
conditions as IRS2. Over the wavelength range of the \brg\ line, we
substituted a spectrum of telluric features in place of the
standard spectrum, instead of interpolating across the \brg\ line of
the standard star as is usually done. As we are interested in the
broad features in the line profile in IRS2, we prefer to have easily
recognizable residuals from narrow telluric lines rather than broad
residuals from the inaccurate removal of the broad \brg\ line of the
standard star.

L-band spectra of both IRS2 and IRS2b were obtained using ISAAC on
February 22, 2002, with a resolving power $R \simeq 1\,200$ .  The
response function was derived by dividing the spectrum of an A6~II
star (HD~73634), observed under similar conditions, by a synthetic
spectrum compiled from a {\sc marcs} model atmosphere with parameters
\teff\ = 8\,100 K, $\log g$ = 2.76, and [Fe/H]=0.00 (L. Decin priv.
comm.). The model assumes plane-parallel layers and solar
abundances. For the synthetic spectrum calculations the atomic
line list of J. Sauval (Decin 2000) and the H-line list of Kurucz
(1991) were used. For more details about the atmosphere models, see
Decin et al. (2000) and Plez et al. (1992).

   \begin{figure}[t]
   \begin{center}  
      \resizebox{8.5cm}{!}{\includegraphics{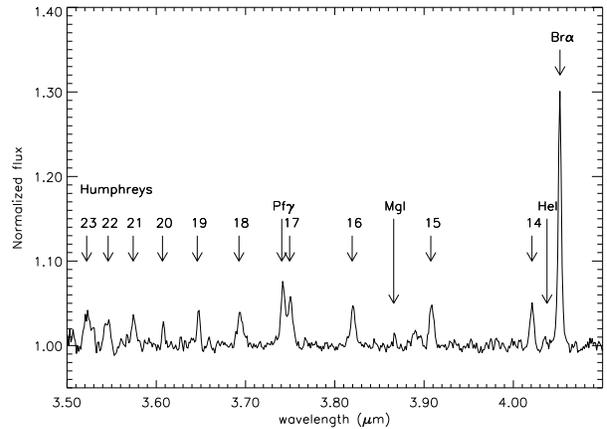}}
      \caption{3.5-4.1 \mum\ spectrum of NGC\,2024 IRS2 obtained
               with CGS4/UKIRT
               at a resolving power of $R \sim 1\,500$. Line
	       identifications 
               and properties are given in Table~\ref{tab_lines}.}
      \label{fig_L}
   \end{center}
   \end{figure}

The L'-band (3.5-4.1\,\mum) spectrum was obtained on December 23, 2000
(UT), using the Cooled Grating Spectrometer 4 (CGS4; Mountain et
al. 1990) on UKIRT.
CGS4's 40 l/mm grating was used in first order with the 300
mm focal length camera and the 0.6\arcsec\ wide slit, giving a nominal
resolving power R$\sim$1500. The array was stepped to provide two data
points per resolution element. A signal-to-noise ratio of $\sim$ 100
was achieved on the continuum. We used the
Starlink Figaro package for data reduction. 
Wavelength calibration was done using the
second order spectrum of an argon arc lamp. The spectrum was reduced
using the F0\,V standard BS\,1474 and the spectrum of a star of the same
spectral type observed at similar spectral resolution with the
Short Wavelength Spectrometer (SWS, de Graauw et al. 1996) on the {\em
Infrared Space Observatory} (ISO, Kessler et al. 1996). We built a
responsivity curve by dividing the L'-band spectrum of the standard star by
the ISO spectrum of this F0\,V star and used it to reduce the
spectrum of IRS2. 

  \subsection{Line properties and identifications}
  \label{sect_line}

\begin{figure}[t!]
\begin{center}
\resizebox{8.5cm}{!}{\includegraphics{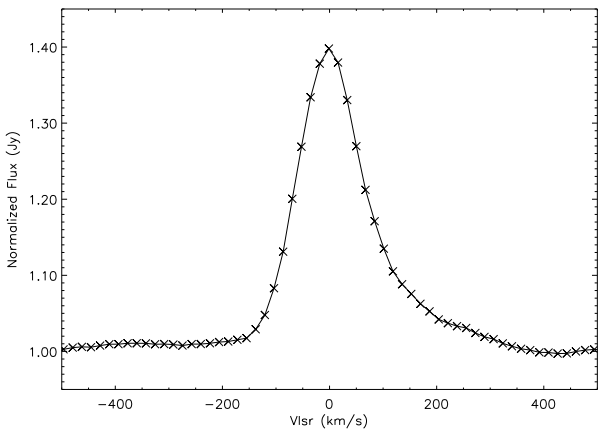}}
\caption{The 2.17 \mum\ \brg\ line of IRS2 obtained with VLT/ISAAC.
         The line peaks at $\vlsr = 0$ \kmsec and has a 
         wing extending $\sim$ 400 \kmsec\ to 
         the red side of the profile.
        }
\label{fig_brg}
\end{center}
\end{figure}  

\begin{table*}[ht!]
\caption{Wavelength, emission line equivalent width, full width at 
         half maximum, and identification of the lines 
         present in the 2.08-2.18 \mum\ and the 3.5-4.1 \mum\ spectra.
         The last columns list the spectral resolving power and beam
	 size.  Lines for which no measured quantities are given
	 represent uncertain identifications.
         \label{tab_lines}}
\begin{center}
\begin{tabular}{lcclcc}
\lam    &EW                   & FWHM   & Identification & $R$ & beam\\
(\mum)  &(\AA)                 &(\kmsec)&                &  &(\arcsec) \\
\hline
 2.090  & 0.30  $\pm$  0.06    &  117   &\feii\,($3d^{6}4p-3d^{6}4s$)& 8000 &0.6 $\times$ 0.9\\
 2.138  & 0.6   $\pm$  0.1     &  290   &\mgii\,($5s-5p$)            & 8000 &0.6 $\times$ 0.9\\
 2.144  & 0.4   $\pm$  0.1     &  320   &\mgii\,($5s-5p$)            & 8000 &0.6 $\times$ 0.9\\
 2.166  & 5.3   $\pm$  0.1     &  150   &\hi\,($4-7  $)              & 8000 &0.6 $\times$ 0.9\\
 3.575  & 1.7   $\pm$  0.3     &  389   &\hi\,($6-21 $)              & 1500 &0.6 $\times$ 1.8\\
 3.608  & 1.3   $\pm$  0.3     &  289   &\hi\,($6-20 $)              & 1500 &0.6 $\times$ 1.8\\
 3.647  & 2.0   $\pm$  0.3     &  337   &\hi\,($6-19 $)              & 1500 &0.6 $\times$ 1.8\\
 3.694  & 2.3   $\pm$  0.3     &  393   &\hi\,($6-18 $)              & 1500 &0.6 $\times$ 1.8\\
 3.742  & 5.0   $\pm$  0.3     &  470   &\hi\,($5-8  $)              & 1500 &0.6 $\times$ 1.8\\
 3.750  & 3.0   $\pm$  0.3     &  470   &\hi\,($6-17 $)              & 1500 &0.6 $\times$ 1.8\\
 3.820  & 2.9   $\pm$  0.3     &  460   &\hi\,($6-16 $)              & 1500 &0.6 $\times$ 1.8\\
 3.866  &       $\le$  0.3     &        &\mgi\,($3s.4f-3s.5g$)       & 1500 &0.6 $\times$ 1.8\\
 3.908  & 3.3   $\pm$  0.3     &  470   &\hi\,($6-15 $)              & 1500 &0.6 $\times$ 1.8\\
 4.021  & 3.5   $\pm$  0.3     &  440   &\hi\,($6-14 $)              & 1500 &0.6 $\times$ 1.8\\
 4.038  &       $\le$  0.3     &        &\hei\,($1s.4d-1s.5f$)       & 1500 &0.6 $\times$ 1.8\\
 4.052  &14.5   $\pm$  0.2     &  290   &\hi\,($4-5  $)              & 1500 &0.6 $\times$ 1.8\\
\end{tabular}
\end{center}
\end{table*}

The 2.08-2.18 \mum\ and 3.5-4.1 \mum\ spectra of IRS2 are presented in
Figs.~\ref{fig_K} and \ref{fig_L}. The spectra only include emission lines,
mainly from hydrogen. We identified two lines of \mgii\ and
possibly one line of \feii\ at \lam 2.089 \mum, one line of \mgi\ at
\lam 3.866 \mum\ and one line of \hei\ at \lam 4.038 \mum. The
\mgii\ and \feii\ lines indicate the presence of very high density and 
fairly cool gas close to the star (e.g. Hamann \& Simon 1987).  
The 2.08-2.18 \mum\ spectrum of IRS2 is similar to the spectra of 
Group 3 Be stars as proposed by Clark \& Steele (2000), having spectral 
types in the range B1--B4. That group contains five stars all showing 
\mgii\ and \brg\ in emission and no \hei; four of them also show \feii\ 
in emission. The He I 2.06 \mum\ line, usually the strongest line in 
the K-band, is not covered in our spectrum, but is not present in the 
K-band spectrum of IRS2 obtained by Thompson et al. (1981).  The
equivalent width (EW) of the \brg\ line is larger than 8 \AA\ in all
Group 3 stars, but is somewhat less in IRS2 (5.3 $\pm$ 0.1 \AA).

The \mgii, \pfg\ and Humphreys series lines are fairly broad with 
full widths at half maximum (FWHMs) of about 400 \kmsec. The Brackett
lines are much narrower, with deconvolved FWHMs of 145 \kmsec\ for
\brg\ and 200 \kmsec\ for \bra.  The values in Table~\ref{tab_lines} 
are not deconvolved by the instrumental profile, but deconvolution has 
only a small effect except for \bra. The difference in FWHM between 
the Brackett and other hydrogen lines is investigated further in 
Sect.\ref{sec_diag}.

Previous near-infrared spectroscopic observations of recombination
lines in NGC2024 IRS2 have been reported by Thompson et al. (1981),
Smith et al. (1984), Chalabaev \& L\'ena (1986), Geballe et al.
(1987a), Maihara et al. (1990), and Nisini et al. (1994).  Both the
line fluxes and line profiles have shown significant variations. Only
small variations in the line fluxes were observed between 1980 and
1989. In 1991, however, the line fluxes increase greatly (up to 10 times
for \bra; Nisini et al. 1994). The proportional increase was larger for
\bra\ than for \brg\ and \pfg. The increase was followed by a decrease
during the period 1992--1994. There have been no reported measurements 
of these lines since 1994.

High-resolution spectroscopy ($R \sim$ 10\,000) by Smith et al. (1984)
and Chalabaev \& L\'ena (1986) showed that the peaks of the \bra, \pfg,
and \brg\ lines were at velocities in the range $20 < \vlsr < 50$
\kmsec\ and that the lines had extended blue-shifted wings. In 1992 all
three lines peaked at \vlsr\ = -25 \kmsec\ and had extended red-shifted
wings (Nisini et al. 1994). Our 2000 spectrum shows the peak of \brg\
near \vlsr\ = 0 \kmsec\ and a very extended wing on the red side of the
profile (see Fig.~\ref{fig_brg}). The widths of the lines, both at half
maximum and at zero intensity (FWZI), are comparable in 1984, 1992 and
1993 and are about 100 and 250 \kmsec, respectively. However, in 2000
all three lines had much larger widths. The measured FWZI of $\sim$ 600
\kmsec\ for the \brg\ line is much larger than has previously been seen
in any of the lines. The large width is mainly due to the red wing.

   \begin{figure*}[t!]
    \resizebox{17cm}{!}{\includegraphics{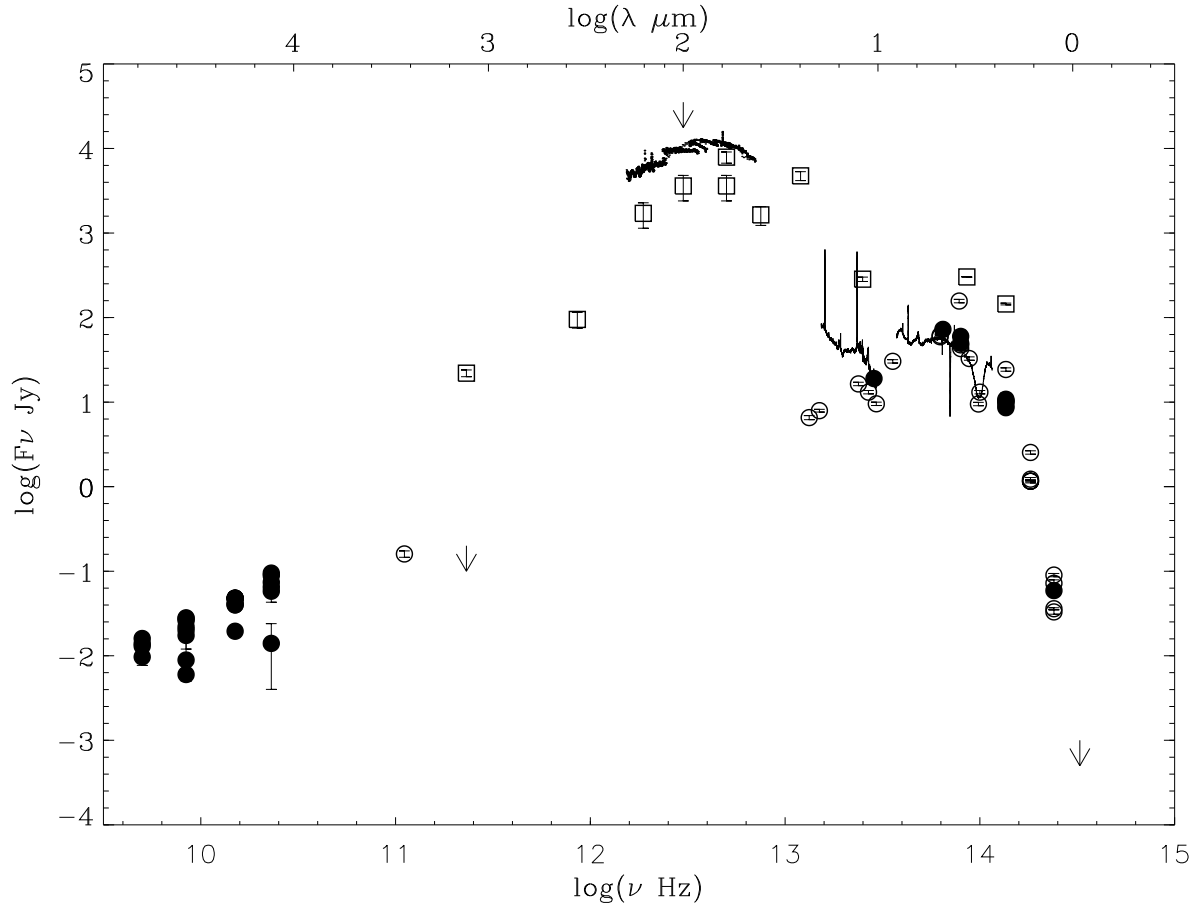}}
    \caption{The spectral energy distribution of IRS2, compiled from
    published photometric data and ISO SWS/LWS spectra.  Details
    concerning these data are given in Tables~\ref{tab_short},
    \ref{tab_short_cm}, and \ref{tab_long}. The high resolution ISO
    spectra (shown as continuous lines) are obtained in AOT6
    mode. Open squares denote measurements made with apertures larger
    than 40\arcsec; open circles indicate aperture sizes of
    5--20\arcsec; closed circles represent beam sizes less than
    5\arcsec. Most often, the error bars on the data are smaller than
    the symbol size. The subdivision in aperture size clearly shows
    that the emission bump peaking at about 100 \mum\ is not due to
    the nearby circumstellar medium of IRS2, i.e. it is not associated
    with the system itself but must come from warm material in the
    vicinity of the star.
            }
             \label{fig_sed}
   \end{figure*}

Both the ISAAC and UKIRT measurements are in smaller apertures than
all previous observations.  The narrower profiles measured by Smith et
al. (1984), Chalabaev \& L\'ena (1986) and Nisini et al. (1994) might
result from a larger contribution from extended gas with lower
velocity dispersion. Our new high-resolution observations demonstrate
that eight years after the outburst the lines have not returned to
their pre-outburst width, shape, and peak velocity. Therefore we
question if \vlsr\ = 40 \kmsec\ is the actual systemic velocity for 
IRS2. We note that the extended molecular cloud of NGC~2024 has \vlsr\ 
of 6 \kmsec\ in the region surrounding our program star (Kr\"ugel et 
al. 1982). IRS2 likely has a similar velocity. We do not pursue on the 
nature of the observed line variability as this requires consecutive 
observations on timescales much shorter than are available. This 
variability was already pointed out by Nisini et al. (1994), who also 
suggest several explanations for the line behaviour between 1981 and 
1994.

\section{The spectral energy distribution}
\label{sect_sed}

In Fig.~\ref{fig_sed} we display the spectral energy distribution
(SED) of IRS2 from the near-infrared to the radio domain. The SED is
based on photometric data compiled from the literature plus spectra
obtained using the SWS and LWS instruments on ISO.  The interpretation
of the plotted data is not straightforward, as measurements have been
obtained with widely different apertures. We provide information on
all data points in the figure, including beam sizes, measured
magnitudes and fluxes, as well as the observational technique (when
non-photometric) in tables~\ref{tab_short}, \ref{tab_short_cm}
and~\ref{tab_long} in appendix.

\subsection{Contribution of IRS2 to the observed spectral
energy distribution} \label{sect_infr}

Variations between different observations at the same wavelength can
be noticed, the most striking being in the range of 3 to 30 \mum. It
is crucial to determine the dominant cause of these variations in
order to measure the flux directly attributable to IRS2.  The
variations could be due to intrinsic variability of IRS2, to the
extended background, or to other discrete sources. The region
surrounding IRS2 is far from empty at infrared wavelengths (cf. Bik et
al. 2003) and we thus expect some dependence of the observed flux on
aperture size. In the following we separately discuss four wavelength
ranges: the near-infrared (1-10 \mum), the mid-infrared (10-25 \mum),
the far-infrared to sub-millimeter (25 \mum\ - 0.2 cm) and the radio
domain (0.2 - 6 cm).

Most of the near-infrared observations were made with beam sizes less
than or equal to 15\arcsec; only the data observed by Frey et
al. (1979) were obtained through a larger aperture (63\arcsec).  The
aperture used by Grasdalen (1974) is not specified, however the
bolometer systems they used on the 1.3m and 2.1m telescopes at Kitt
Peak had aperture diameters less than 12\arcsec\ (D. Joyce
priv. comm.), the ones they used likely being smaller than 8.5\arcsec\
(G. Rieke priv. comm.). The fluxes observed through apertures less
than or equal to 20\arcsec\ (including the ISO/SWS spectrum) do not
show any trend with aperture size.  The higher values obtained by
Nisini et al. (1994), which were observed with simultaneous line
enhancement and shift, most likely originate from variations close to
or in IRS2. The J-band measurement of Grasdalen (1974) is brighter
than other observations by 0.8 mag. It is difficult to attribute this
value to source variability as it is the only band affected by this
discrepancy.  We conclude that IRS2 is the dominant source in the 1-10
\mum\ region through aperture diameters less than or equal to
20\arcsec.

The mid-infrared SED consists of four data sets. Grasdalen (1974) and
Walsh et al. (2001) observed similar fluxes at $\sim$ 10 \mum\ through
small apertures. The ISO/SWS flux is higher than the other mid-IR
values suggesting that IRS2 is not the dominant source in the
mid-infrared within the 14$\times$20\arcsec\ ISO beam. The IRAS fluxes
are 20 and 600 times larger than the ones of Grasdalen (1974) at 12
and 25 \mum\ respectively, and are obtained through much larger
apertures.  Moreover, there is a large offset of 72.8\arcsec\ between
the position of IRS2 and IRAS 05393-0156; this IRAS point source has
commonly been associated with IRS2 (e.g. Kurtz et al.  1994 or Walsh
et al. 1998), but is actually closer to FIR5 (Mezger et al. 1988),
supporting the view that IRS2 is not the dominant source in the IRAS
beam.
    
The IRS2 region has not been observed at far-infrared to
sub-millimeter wavelengths with aperture sizes smaller than 40\arcsec.
ISO/LWS observed three different sources between 43 and 197 \mum\
within NGC\,2024: IRS2, FIR3 and FIR5. The beams including IRS2 and
FIR3 overlap by about 50\% whereas the beam of FIR5 has only a very
small fraction in common with the others. The three spectra are very
similar suggesting that at these wavelengths the emission is coming
from the extended surroundings and not from individual sources (see
Giannini et al. 2000).

In the radio domain we only consider data obtained with
interferometers, i.e. with small beam sizes. After inspection of the
original data, we feel that the 6 cm data point of Snell \& Bally
(1986; 2.8 mJy) is too low and that the 1.3 cm point (Gaume et al.
1992; 19 mJy) suffers from significant uncertainties. Recalibrated
values are given in Table~\ref{tab_short_cm} together with other
reported fluxes and VLA archival data. Data taken between May 1994 and
January 1995 all show similar flux levels. Observations by Kurtz in
1989 and by Walsh in 1994 both show smaller fluxes by about a factor
of three to four.  The flux differences cannot be attributed to
differences in beam size. Radio variability and spectral index is
discussed in Sect~\ref{sec_radio}.

We conclude that the bulk of the mid- and far-infrared emission
observed in large apertures and peaking at $\sim$ 100 \micron\ is {\em
not} associated with the nearby circumstellar medium of IRS2, but
arises from a more diffuse source. Therefore, one cannot use the IRAS
color-color criteria for \uchii\ regions (Wood \& Churchwell 1989) to
conclude that IRS2 is a massive star surrounded by a dust cocoon.
This weakness of the infrared criterion for \uchii\ was already
pointed out by Codella et al.  (1994) and Ramesh \& Sridharan (1997).
Except for the shortest infrared wavelengths, only small aperture
observations ($\leq$ 20\arcsec) are relevant in probing the nature of
IRS2 and its immediate circumstellar environment. Large aperture
measurements will not be considered further in this paper.

\subsection{Extinction}
\label{sec_ext}

Estimates for the extinction towards IRS2 are very uncertain and
range between \AV\ = 12 and 32 mag (Thompson et al. 1981, Grasdalen
1974).  The identification of IRS2b as the ionizing source of the
Flame Nebula, 5\arcsec\ away from IRS2 (Bik et al. 2003), allows for
the first time an independent determination of the interstellar
extinction caused by the elongated structure obscuring the center of
NGC\,2024. Applying the synthetic extinction law of Cardelli et al.
(1989), Bik et al. (2003) find an \AV\ of 24 $\pm$ 0.5 mag towards
IRS2b, taking
R$_{\rm V}$ = 5.5.  Note that the value of R$_{\rm V}$ is relevant for
the absolute value of the extinction, but has only a weak effect on
the slope of the near-infrared spectrum.  IRS2 and IRS2b are most
likely located in the \hii\ region behind the cold obscuring dust bar
and in front of a warm molecular cloud (Barnes et al.  1989).  The
2.08-2.18 \mum\ spectrum of IRS2b does not show any evidence of
circumstellar extinction (Bik et al. 2003), implying that the
obscuration is dominated by the cold foreground molecular cloud.

\begin{figure}[t!]
    \begin{center}         
    \resizebox{8.5cm}{!}{\includegraphics{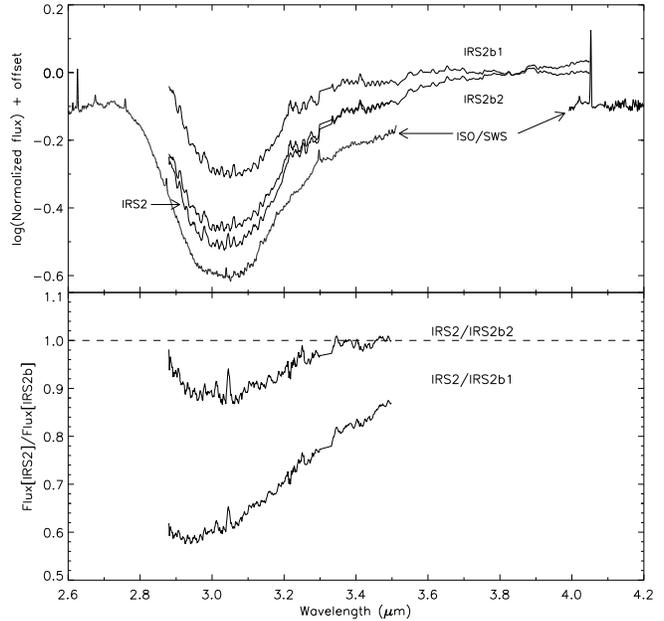}}
    \caption{
             The upper panel shows the 3.0 \mum\ ice band in the normalized
             L-band spectra of IRS2b and IRS2, as well as the
             normalized ISO/SWS spectrum of IRS2 (shifted by
             -0.1). IRS2b is normalized using the slope of the 3.4 to
             3.5 \mum\ region (denoted IRS2b1) and of the 3.5 to 4.0
             \mum\ region (IRS2b2).  The lower panel shows the ratio
             of the ice feature in IRS2 and in IRS2b, indicating that
             the extinction towards IRS2 is higher.}
    \label{fig_ice}        
    \end{center}           
    \end{figure}

To compare the interstellar extinction towards IRS2 and IRS2b, we have
inspected the strength of the 3.0 \mum\ water ice absorption seen
towards both sources. Whittet et al. (1988) and Teixeira \& Emerson
(1999) present evidence for a linear relation between the optical
depth in the ice band and the visual extinction. The parameters of
this relation is expected to depend on the physical and chemical
properties of the intervening medium, and thus may vary from cloud to
cloud.
The main uncertainty in the extinction determination is the
uncertainty in the continuum flux level near the ice feature. The top
panel of Fig.~\ref{fig_ice} shows normalized L-band spectra of both
IRS2 and IRS2b. The spectrum of IRS2 stops at 3.5 \mum\ due to
saturation effects; that of IRS2b extends out to 4.0 \mum. IRS2 has
been observed by ISO/SWS. This allows a reliable determination of the
continuum in our L-band spectrum, as the ISO spectrum covers a much
broader wavelength interval. The normalization of the spectrum of
IRS2b is problematic.  First, the spectrum starts at 2.9 \mum, which
is somewhat inside of the ice feature. Therefore, we cannot use the
short wavelength side of the profile to define the continuum. Second,
just outside of the feature at the long wavelength side the spectral
slope shows a kink at about 3.5 \mum. We consider the two limiting
cases: IRS2b1, in which the continuum is based on a slope measurement
in the 3.4 to 3.5 \mum\ region, and IRS2b2, in which the continuum is
derived from the 3.5 to 4.0 \mum\ interval. The lower panel of
Fig.~\ref{fig_ice} shows the ratio of the L-band spectra of IRS2 and
IRS2b for these two cases.

For both normalizations the ice band absorption in IRS2 is found to be
stronger than in IRS2b, showing that more interstellar ice is present
in the line of sight towards IRS2. Using \AV\ = 24 mag for IRS2b, we
arrive at a propotional \AV\ of 27 (using IRS2b1) to 41 mag (using
IRS2b2) for IRS2. If we require an intrinsic spectral index of two,
the upper limit can be adjusted down to \AV\ = 36.5 mag (for R$_{V}$ =
5.5). We conclude that the visual extinction towards IRS2 is between
27 and 36.5 mag.

   
\section{Diagnostic tools}
\label{sec_tool}

We discuss a number of independent diagnostic methods that may help to
constrain the geometry of the circumstellar material around
IRS2. First, we use the radio slope to determine the gradient in the
radial density structure. We compare this slope with values typical
for spherical outflows and disks. Second, we apply a technique
developed by Lenorzer et al. (2002a) in which the flux ratio of lines
in the L'-band are used to determine whether the material is
distributed in a spherically symmetric or disk-like structure. Third,
we briefly summarize the important observational and modeling findings
made on CO band-head emission about IRS2.

\subsection{Radio slope}
\label{sec_radio}

\begin{figure}[t!]
\begin{center}
\resizebox{8.5cm}{!}{\includegraphics{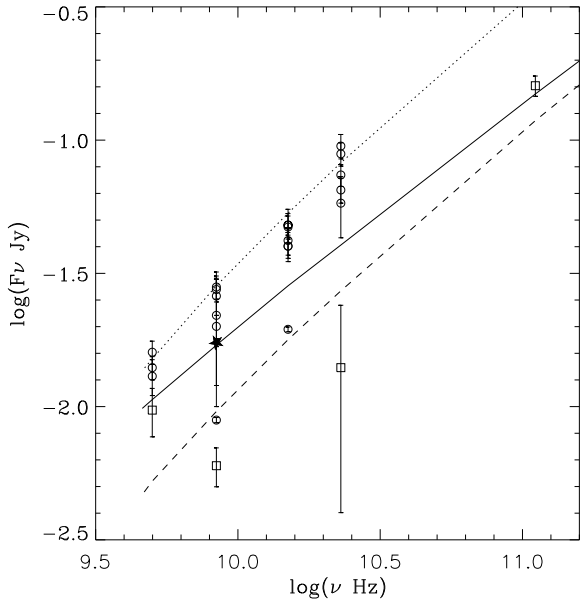}}
\caption{Radio observations of IRS2. Multi-frequency observations at a
         single epoch are denoted by open circles.  Squares represent
         non simultaneous data. Recent observations by Rodr\'\i guez
         (priv. comm.)  resolving the radio source are denoted by a
         filled star.  Assuming a spherical wind, these data are
         fitted by $n_{\circ} = 10^{11}$ cm$^{-3}$ and $\Rout = 145$
         AU.  Data point obtained at other epochs yield roughly
         similar parameters: 1989 observations by Kurtz et al. (1994)
         imply the same $n_{\circ}$ and $\Rout = 25$ AU (dashed line);
         1994 and 1995 data require $n_{\circ} = 2.5 \times 10^{11}$
         cm$^{-3}$ and recombination at 42 AU (dotted line).
        }
\label{fig_radio1}
\end{center}
\end{figure}

The radio observations are listed in Table~\ref{tab_short_cm} (in
appendix) and plotted in Figs.~\ref{fig_radio1} and~\ref{fig_radio2}.
Discarding the problematic points taken in 1981-83 (Snell \& Bally
1986) and 1989 (Gaume et al. 1992) (see Sect.~\ref{sect_sed}), flux
variations of up to about a factor of three are observed on timescales
of years, and variations of roughly a factor of two on timescales of
months. Data prior to 1994 indicate relatively low flux densities,
whereas the 1994-1995 dataset suggests a significant brightening of
the source, followed by a possible relaxation in 2002.  The nature of
this variability is not understood, however it may suggest some sort
of flaring event.
The spectral indices $\alpha \equiv \partial \log \Fnu / \partial \log
\nu$ derived from simultaneous measurements suffer from large
uncertainties, but do not show a trend in time, nor do they depend on
frequency base. Simultaneous observations yield spectral indices
ranging from 0.75 to 1.33, with a weighted average of $\alpha$=1.25
$\pm$ 0.06.

The radio spectral index can be used as a diagnostic of the density
structure in a circumstellar medium.  For a spherical, isothermal
stellar wind that has reached its terminal velocity it can be shown
(Wright \& Barlow 1975; Panagia \& Felli 1975; Olnon 1975) that
$\alpha = 2/3$, neglecting a very modest dependence of the Gaunt\,{\sc
iii} factor on frequency.  The observed slopes are somewhat
steeper, suggesting either that the ionized region is sharply bounded
(Simon et al. 1983), or that the radial density distribution $\rho(r) 
\propto r^{-m}$ has a gradient steeper than that for a constant velocity
wind, $m=2$. For the wind model described in
Sect.~\ref{sec_discussion} with $m=2$, two free parameters essentially
determine the radio spectrum: the electron density at the stellar surface
and the recombining radius.  Recent observations by L. F. Rodr\'\i guez
(priv. comm.) have resolved the radio source, its diameter 
is 0.8 \arcsec\ at 3.6\,cm, corresponding to a radius of 145 AU.  
This outer radius and the
flux measured by Rodr\'\i guez require an electron density of 
$n_{\circ} = 10^{11}$ cm$^{-3}$ (solid line in Fig.~\ref{fig_radio1}).
Data obtained at different epochs are matched by roughly similar 
combinations of outer radius and $n_{\circ}$ (see also 
Fig.~\ref{fig_radio1}).

\begin{figure}[t!]
\begin{center}
\resizebox{8.5cm}{!}{\includegraphics{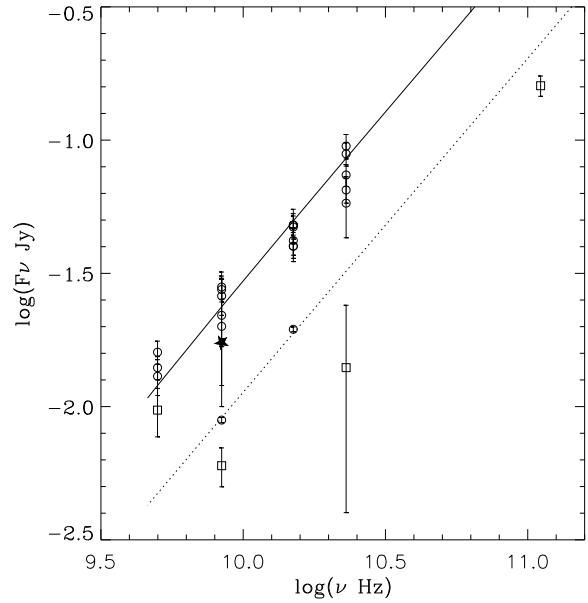}}
\caption{
         Same data as in Fig~\ref{fig_radio1}. Here the overplotted
         models are for disks. 
         We fit the radio observations of IRS2 with a disk
         model. Both models are for a slope of $m=3.17$ in the 
         radial density distribution. The solid (dashed) line assumes
         an electron density at the stellar surface of 6.7 (1.8)
         $\times$ 10$^{16}$ cm$^{-3}$ and an outer radius of 77
         (39) AU, respectively, to prevent a turn-down of the radio spectrum.
        }
\label{fig_radio2}
\end{center}
\end{figure}

In the more general case that a spherically symmetric density
distribution is given by a power-law, the spectral index is given by
$\alpha(m) = (4m-6)/(2m-1)$ where again the Gaunt\,{\sc iii}
dependence on frequency is neglected. To match the observations one
finds $m= 3.17$ for $\alpha = 1.25$.  The $\alpha(m)$ relation also
holds when the gas is in a disk with a constant opening angle, if the
density still has a symmetric radial dependence.  Disk models applied
to different epochs are presented in Fig.~\ref{fig_radio2} for an
opening angle of 5 degrees. The free parameters of these models are
(again) the electron density at the stellar surface and the disk
radius.  The electron density required to match the flux density is
high (between 1.8 and 6.7 $\times$ 10$^{16}$ cm$^{-3}$ at the stellar
surface).
Waters et al. (1991) observed six classical Be-stars at millimeter
wavelengths and found their spectral indices to be in the range $0.94
\le \alpha \le 1.45$. These values are comparable to the ones for
IRS2. However, typical densities for these ``classical disks'' are
several orders of magnitude lower than what is required for IRS2.

The present data do not allow us to conclude whether the observed
radio variability is due to changes in the ionization and/or density
structure. In Sect.~\ref{sec_discussion} we will use the overall SED
to further constrain the circumstellar environment.

\subsection{Infrared hydrogen recombination lines}
\label{sec_diag}

\begin{figure}[t!]
\begin{center}
\resizebox{8.5cm}{!}{\includegraphics{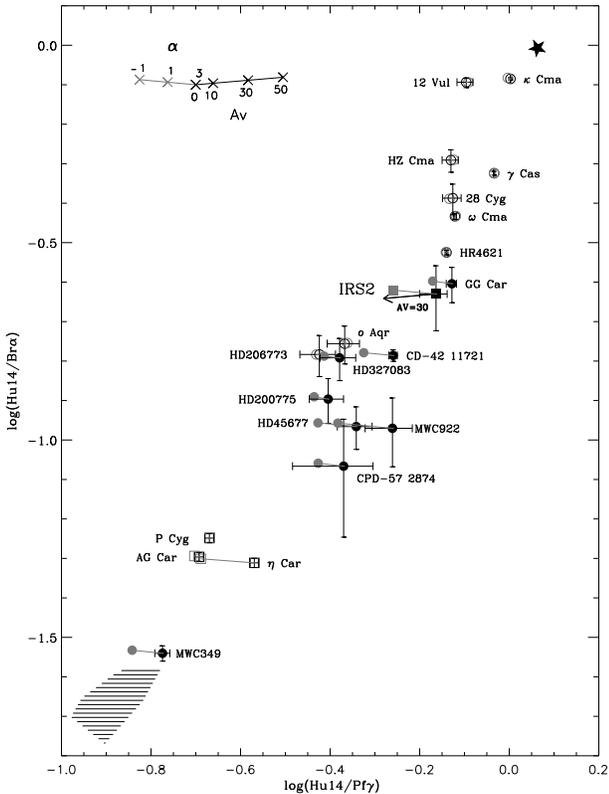}}
\caption{\hu/\bra\ versus \hu/\pfg\ for hot stars observed with ISO
         (Lenorzer et al. 2002b). The different classes of objects,
         LBVs (squares), B[e] stars (filled circles) and Be stars
         (open circles) are well separated.  IRS2 is denoted by a
         filled square. The filled asterisk indicates the position of
         optically thick black-body emission; the striped region shows
         the range for Menzel case B recombination, including
         collisional de-excitation, for temperatures higher than
         $10^{4}$ K. Line flux ratios are plotted in black; EW ratios
         in grey.  \label{diag}
        }
\end{center}
\end{figure}

Lenorzer et al. (2002a) recently developed a tool to investigate the
geometry of the circumstellar material around hot stars.  A diagram
using hydrogen lines from different series, all observable in the
L'-band, allows one to separate sources in which the circumstellar
material is distributed roughly spherically (e.g. LBV stars) from
sources in which it is more disk-like (e.g. Be stars). The diagram
(Fig.~\ref{diag}) is based on 21 spectra obtained by ISO/SWS (Lenorzer
et al. 2002b).  It shows that both line flux ratios \hu/\pfg\ and
\hu/\bra\ increase from LBV to B[e] to Be stars. This trend can be
understood in terms of the span in absorption coefficients of these
lines.  In an optically thin medium one expects the flux ratios to
follow Menzel Case B recombination theory. At the other extreme, in an
optically thick medium, the ratios become independent of the
absorption coefficient as the flux in any line is determined by the
size of the emitting surface.  Some limitations of this diagnostic
tool are to be expected, as the correlation between geometry and
optical depth is established from objects for which the geometry of
the circumstellar material is reasonably simple. Deriving the geometry
of objects that might have multi-components and vary on short
timescales is not trivial and requires additional diagnostics,
e.g. the modeling of the line profiles.

To investigate the effect of extinction the diagram includes the line
flux and EW ratios. The EW diagram is calibrated on the line flux
diagram such that P~Cygni, which has negligible extinction, is at the
same location in both diagrams.  Fig.~\ref{diag} shows that IRS2 falls
in the upper right part of the diagram, close to the B[e] star
GG\,Car. This position indicates that its lines are predominantly
formed in optically thick gas. This situation likely arises in a
disk-like geometry. The position of the line flux ratio of IRS2
corrected for an \AV\ of 30 mag (Sect.~\ref{sec_ext}) is also
indicated in the figure. There is a close agreement between the
de-reddened line flux ratio and the EW ratio of IRS2.  The large shift
observed for B[e] stars is not predominantly caused by extinction (all
plotted sources have very moderate \AV), but by a rising dust
continuum reducing their spectral indices. We conclude that the
continuum of IRS2 is not affected by a dust continuum strongly rising
between 3.74 and 4.05 \mum.

To further constrain the geometry of the circumstellar material, we
investigate the line widths.  The Humphreys lines as well as the \pfg\
line have FWHMs of about 400 \kmsec; the S/N of the L'-band spectrum
is too low to derive any trend in the widths of these lines. The
Brackett lines are narrower with FWHMs of about 200 \kmsec\
(Sect.~\ref{sect_line}), suggesting a different line forming region
for intrinsically stronger lines.

The ratio of the FWHMs of \pfg\ and \bra\ is about 2.0. For typical Be
stars like $\gamma$~Cas, HR4621, $\kappa$~Cma or in the B[e] star
GG~Car, close to the position of IRS2 in the diagram, this ratio lies
between 1.1 and 1.3, whereas it is less than 1.0 for wind sources like
P~Cyg. The fact that the Brackett lines are much narrower than the
other hydrogen lines suggests that they have a significant
contribution from low density gas close enough to the star to be
observed in our narrow-slit spectrum. The FWHM of this optically thin
component can be estimated from the high-resolution spectrum of the
\brg\ line as $\le$ 130 \kmsec.  The infrared recombination lines
observed in IRS2 are mainly produced in a high velocity (400 \kmsec)
optically thick medium, with a small contribution from a lower
velocity ($\le$ 130 \kmsec) optically thin medium. The high velocity
contribution in itself would give IRS2 a position in Fig.~\ref{diag}
that is closer to the Be stars.

\subsection{CO band-heads}
\label{CO}

CO band-head emission at 2.3 \mum\ was first detected in
NGC\,2024/IRS2 by Geballe et al. (1987b). Its presence attests of the
existence of high-density and warm molecular gas. CO is largely
dissociated at T $\ga$ 5\,000 K, while for T $\la$ 2\,000 K the upper
vibrational levels will not be excited frequently by collisions.  In
addition to the proper temperature range, a density of at least
10$^{10}$ cm$^{-3}$ is required to maintain the population of the
excited vibrational levels, which radiatively decay rapidly via
fundamental band transitions.  The 2-0, 3-1, 4-2 and 5-3 band-heads
were observed by Geballe et al. at a resolution of $\sim$ 460 \kmsec,
insufficient to resolve the individual CO lines.  >From the relative
band-head intensities they estimated an excitation temperature of
about 4\,000 K. From the intensity of the 2-0 band-head
they found a minimum area of the CO line emitting region of 2.8
$\times$ 10$^{24}$ cm$^2$, assuming that the lines close to the
band-heads are optically thick.

Chandler et al. (1993, 1995) observed the 2-0 band-head at high
spectral resolution and proposed two models to explain the
observations: an accretion disk and a neutral wind. The accretion disk
model requires an extinction in the K-band of about 4.4 mag,
corresponding to \AV\ $\sim$ 33.0 mag (for R$_{\rm V}$ = 5.5). The
best fit values in the case of a disk are: an inclination angle of 33
degrees (i.e. rather face on), and $\mdot_{\rm acc}=5~\times 10^{-7}$
\msunyr. Assuming a stellar mass of 15 \Msun\, the emission arises
between 60 and 180 \Rsun\ where Keplerian velocities are 130-220
\kmsec. The wind model requires a mass-loss rate of about 10$^{-5}$
\msunyr\ and maximum velocities of 100 to 500 \kmsec.

\section{Discussion}
\label{sec_discussion}

In this section we discuss three simple configurations that may
reproduce the observed SED and investigate their relevance based on
the diagnostic tools presented in Sect~\ref{sec_tool}. We evaluate the
contribution of emission from an ionised wind and disk, as well as
that of dust. We end the section with a discussion on the nature of
IRS2.

For the emission from ionized gas we use a basic disk model,
consisting of isothermal hydrogen in Local Thermodynamic Equilibrium
(LTE) and consider bound-free and free-free emission.  We assume the
density distribution to be a power-law, i.e.
\begin{equation}
   \rho(r) = \rho_{\circ} \left( r/r_{\circ} \right)^{-m} \hspace{5mm}
   {\rm or} \hspace{5mm} n(r) = n_{\circ} \left( r/r_{\circ}
   \right)^{-m}
\end{equation} 
within the disk, where the density $\rho$ and the electron density $n$
have the value $\rho_{\circ}$ and $n_{\circ}$ at the stellar surface.
The disk has a constant opening angle $\theta$. The total bound-free
and free-free optical depth for a beam passing the disk in the
direction perpendicular to the mid-plane is given by
\begin{equation}
   \tau^{\rm bf+ff}_{\nu}(T,r) =2 \, \kappa^{\rm bf+ff}_{\nu}(T) \,
   \int_{0}^{{\rm r \times tan}({{\theta}\over{2}})} \rho(r)^{2}\,dz
\end{equation}
It follows that in the part of the spectrum produced by a partially
optically thick medium, the spectral index is given by $\alpha =
(4m-6)/(2m-1)$ (see Waters 1986).

We adopt an opening angle of $\theta =5$ degrees, consistent with the
small opening angles derived from spectropolarimetric observations of
gaseous disks around classical Be stars (e.g. Putman et al. 1996).
Note that this parameter influences the flux level similarly to
\rhozero. The adopted gas temperature is 10\,000 K. The free
parameters of this model are then the density at the stellar surface
\rhozero, the density gradient index $m$, and the outer radius of the
disk \rdisk. The density parameter $\rho_{\circ}$ fixes the total flux
while $m$ determines the spectral index.  The outer radius sets the
wavelength at which the break occurs in the spectral index.
The spherical wind model is very similar to the disk model, except for
the difference in geometry, and the use of a fixed density gradient
$m=2$. 

We remark that for a dense disk the assumption of an isothermal
medium is likely to break down, leading to lower temperatures in
the midplane region as a result of self-shielding. This is not
expected to change the predicted disk energy distribution
significantly as this is set by emission from the disk surface
layers, which can be illuminated directly and therefore have a high 
temperature. 

%
Finally, dust emission, which cannot arise in the ionized
region, is modeled as black-body emission at temperature \tdisk.

There is very little known about the stellar content of IRS2.  From
radio observations of the compact radio source associated with IRS2,
Kurtz et al. (1994) derive an ionizing source corresponding to a B3\,V
star. K-band spectroscopy is consistent with an early B star. In the
following, we assume the central object is a single early-B
main-sequence star.

\begin{figure*}[ht!] \begin{center}
\resizebox{15cm}{!}{\includegraphics{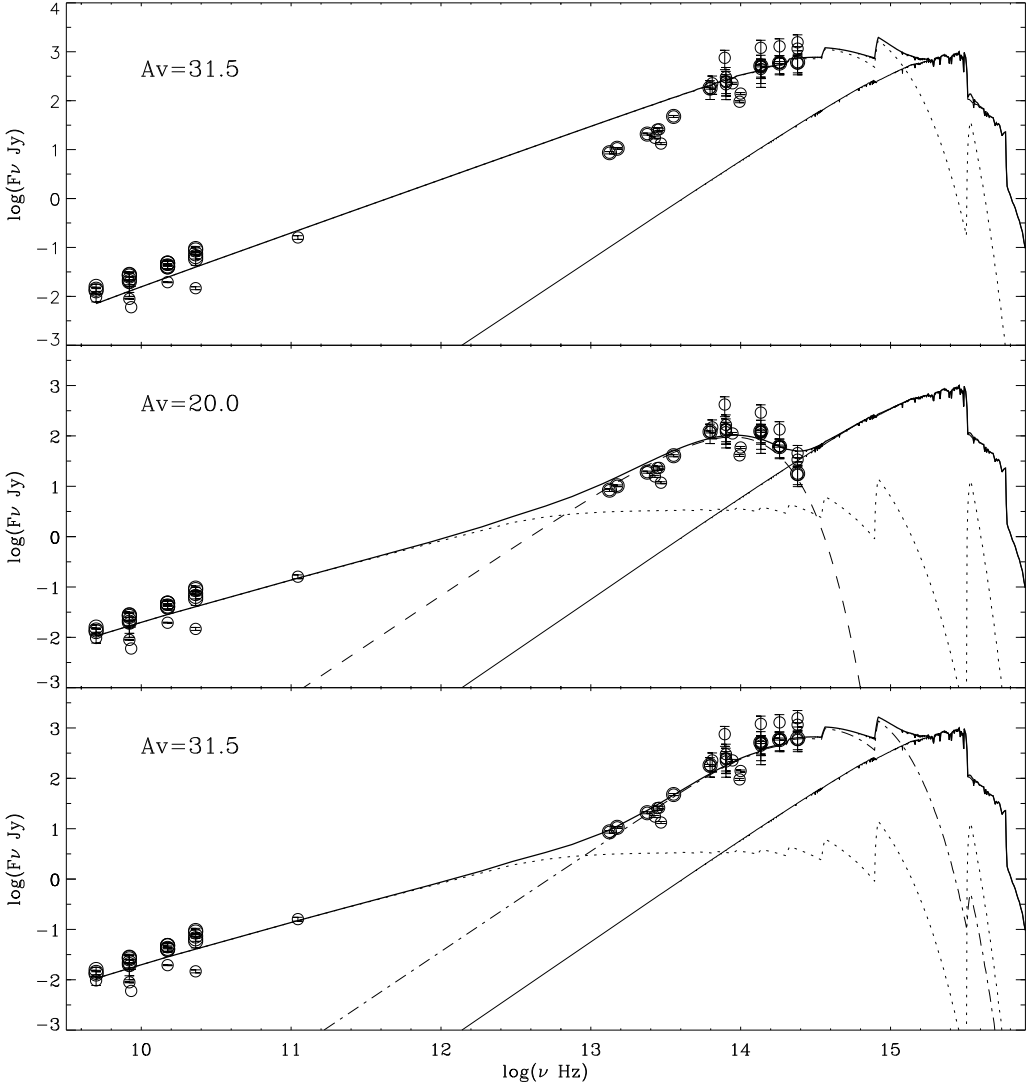}} 
\caption{
         Three models describing the circumstellar medium of IRS2. The
         central star is represented by a Kurucz model of $\teff =
         33\,000$ K and $\log(L/\Lsun) = 4.87$ (solid line). The
         observations are denoted by circles Note that the
         observations are de-reddened using a lower extinction in the
         middle panel (\AV\ = 20 instead of 31.5 mag). From top to
         bottom: Model~A: The dotted line is the disk model and the
         thick solid line is the sum of the stellar and disk
         contributions; Model~B: The dotted line is the wind model,
         the dashed line is the dust model and the thick solid line is
         the sum of the stellar, dust and wind contributions; Model~C:
         The dotted line is the wind model, the dashed line is the
         disk model and the thick solid line is the sum of the
         stellar, disk and wind contributions.  \label{fig_model}
        }
\end{center}
\end{figure*}

\subsection{Model A: a gaseous disk}
\label{sec_modelA}

We first consider the case in which the entire spectral energy
distribution is produced by one gaseous component (i.e., no
dust). This model requires an interstellar extinction high enough to
account for the sharp turn down observed at near-infrared
wavelengths. This is achieved for \AV\ $\ge$ 31.5 mag. After
dereddening, the best fit to all data points yields a spectral slope
$\alpha = 1.09$, corresponding to a density gradient $m=2.7$. The
steep gradient favors a disk-like geometry. Matching the overall flux
level requires an electron density $n_{\circ} = 10^{15}$ cm$^{-3}$.
The radio data do not show a break in the spectral index down to 6 cm,
implying an outer radius of at least 100 AU.  The best fit model is
plotted in the top panel of Fig.~\ref{fig_model}. The model
corresponds to a hydrogen disk mass of $M_{\rm disk}$=3.0 $\times$
10$^{-6}$ \Msun\ and a disk luminosity $\log$($L_{\rm
disk}$/\Lsun)\,=\,4.6.

To investigate if such a very dense gaseous disk allows the excitation
of CO band-heads, as is observed, we determine self-consistently the
temperature throughout the disk using a technique developed by Millar
\& Marlborough (1998). This method adopts a Poeckert-Marlborough disk
model assuming hydrostatic equilibrium.  The continuity equation fixes
the radial density gradient (see Marlborough 1969 and Poeckert \&
Marlborough 1978).  The state of the gas is solved subject to
statistical and radiative equilibrium. The final result is a set of
self-consistent temperatures.  For additional details of the
temperature calculation we refer to Millar \& Marlborough (1998,
1999a, 1999b) and Millar, Sigut, \& Marlborough (2000).

\begin{figure}[t!]
   \begin{center}
   \resizebox{8.8cm}{!}{\includegraphics*[20mm,140mm][195mm,206mm]{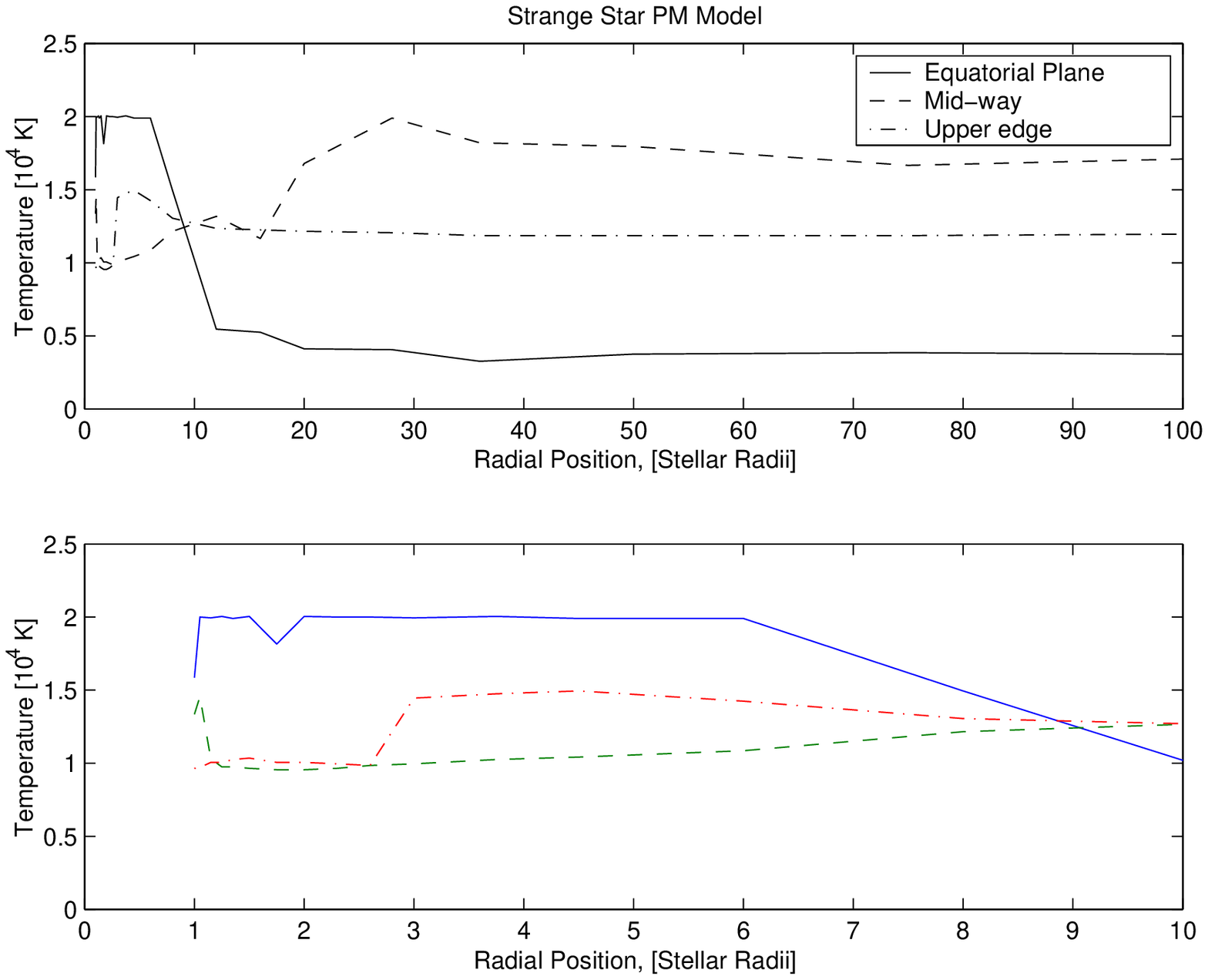}}

   \caption{The temperature structure for a Poeckert-Marlborough disk
            with $n_{\circ} = 3.33\,\times 10^{14}$ cm$^{-3}$ in the
            equatorial plane at the stellar surface. Temperature
            profiles for three different heights above the equatorial
            plane are given. This calculation shows that CO band-head
            emission may arise in a dense gaseous disk.
            \label{grayplot}
        }
   \end{center}
\end{figure}

The adopted parameters are $n_{\circ} = 3.33\,\times 10^{14}$
cm$^{-3}$ in the equatorial plane at the stellar surface and a scale
height rising from 0.1 \Rstar\ at the stellar surface to 34.75 \Rstar\
at a radial distance of 100 \Rstar. Note that this disk structure is
somewhat different from that of our simple model, but close enough for
the qualitative purposes.
Fig.~\ref{grayplot} shows the temperature structure calculated for a
central star having \teff\ = 25\,000 K, \Rstar\ = 10 \Rsun, and $\log
g_{\star} = 3.67$ ( representative of a B1\,V star).  The
global-average kinetic temperature of the disk gas is 13\,700 $\pm$
4\,000 K.  The temperature at the surface of the disk decreases from
15\,000 K close to the star to 12\,000 K over most of the
surface. Mid-way between disk surface and equatorial plane the
temperature rises above that of the surface. At radial distances
beyond 15 \Rstar\ the temperature reaches approximately 17\,000 K. In
the equatorial plane however, the temperature drops drastically at
around 10 \Rstar\ from 20\,000 to 5\,000 K, decreasing outwards to
3\,000--4\,000 K. This is due to a large optical depth for
photo-ionizing stellar radiation. Note that our detailed temperature
computation assumes a pure hydrogen disk. Additional and more
efficient cooling from metals and molecules at these temperatures
would likely result in even lower temperatures. It is therefore
plausible that the observed CO band-heads, representative of a 4\,000
K medium, originate from inside a {\em dense} gaseous disk.  The
observed CO emission does not necessarily require the molecules to be
shielded by a dusty medium.
The position of IRS2 in the line-flux ratio diagram (see
Fig.~\ref{diag}) can be explained qualitatively with a massive Be-like
disk, the shape of the lines, however, require a more complex
geometry.

Model~A is fairly compatible with the diagnostic tools discussed in
Sect.~\ref{sec_tool}, namely the observed radio slope, the infrared
hydrogen recombination lines and CO band-heads, as well as with the
extinction determination. The main shortcoming of this model is that
it overestimates the observed mid-infrared flux in the 8-20 micron
range.  Also, the disk luminosity is a sizeable fraction of the
luminosity of a main sequence star (about 50\,\%).  For an active
disk, i.e. a disk that has an extra source of heating, (e.g. from
accretion) this fraction would decrease.  To get a rough idea, for a
star of 15 \Msun\ and an accretion radius of 10 \Rsun, the disk
luminosity corresponds to an Eddington accretion rate of the order of
10$^{-3}$ \msunyr.  Such accretion rates may perhaps occur in hot
molecular cores, but seem unrealistically high for IRS2.

\subsection{Model B: a stellar wind and optically thin dust}
\label{MODELB}

In the model discussed above, we assumed that one component is
responsible for the entire SED. As can be seen from
Fig.~\ref{fig_model}, Model~A overestimates the mid-infrared
flux. This wavelength range is also observed to have a spectral index
of about 2, different from the radio range. In Model~B we ascribe much
of the infrared continuum emission to a different origin than the
radio emission, i.e. we assume the radio flux to originate in a dense
ionization bounded wind, while a dust shell gives rise to the near-
and mid-infrared flux.

In the middle panel of Fig.~\ref{fig_model}, we show a wind model with
an electron density at the stellar surface of $n_{\circ} = 10^{11}$
cm$^{-3}$ and outer radius of 145 AU by a dotted line
(cf. Sect.~\ref{sec_radio}).  The mass of hydrogen contained in this
wind is $M_{\rm wind} =7.6\,\times 10^{-7}$ \Msun\ yielding a
mass-loss rate of about $\mdot = 1.1\,\times 10^{-6}$ \msunyr\
(adopting a terminal wind velocity of 1000 \kmsec). This mass-loss
rate is about four times larger than predictions for early-B
main-sequence stars by Vink et al. (2001). Assuming 
Baker \& Menzel (1938) Case~B, we find that the above wind density 
structure and outer (recombination) radius are appropriate for
a B1\,V star adopting ionizing fluxes by Panagia (1973).

We model the near-infrared part of the SED by a simple black-body.
Strictly speaking, this is only allowed if the dusty medium
is optically thick. However, for dust temperatures near 1\,500 K
(which are required; see below) dust extinction will be dominated
by species that have a rather grey extinction behaviour at 
near-infrared wavelengths, such as carbon grains. In that case,
the assumption of black-body emission also holds for an optically
thin, isothermal dusty medium.
The monotonic slope of the SED in the 5-20 \mum\ region is quite
insensitive to extinction and shows that if dust is to contribute, it
must have a temperature higher than 1000 K. The highest dust
evaporation temperatures are around 1\,500 K and result in a peak flux
density at 3.4 \micron. To have the flux peak at this wavelength
requires an \AV\ of 20 mag; cooler dust requires less extinction.
This extinction is less than the derived minimum extinction towards
IRS2 of 27 mag. Any contribution to the extinction by circumstellar
dust only aggravates the problem. To minimize the extinction problem
we model the dust emission using a blackbody of 1\,500 K (see dashed
line in middle panel of Fig.~\ref{fig_model}).  The properties of 
the central star require this dust to be at about 240 stellar radii. 
The observed flux density at near-infrared wavelengths then implies
that this dust is optically thin.
The single temperature blackbody is
slightly broader than the observed emission, a problem that cannot be
solved using a temperature distribution to represent the dust as that
would only broaden the emission further.  A higher extinction (than
\AV\ = 20 mag) allows a broader intrinsic near- and mid-IR flux
distribution. However, it would simultaneously shift the peak emission
towards shorter wavelengths.  This would require dust at temperatures
in excess of 1\,500 K, which is not physical.

The near-infrared hydrogen recombination lines could, in principle,
originate in a recombining wind with parameters similar to those
derived above and an \AV\ of 20 mag (Maihara et al. 1990).  The main
problem of model B is that it is incompatible with the minimum
extinction of 27 mag towards IRS2. For this reason the optically thin
hot dust model proposed by Jiang et al. (1984) cannot be supported.

\subsection{Model C: a wind plus a gaseous disk}
\label{sec_modelC}

Finally, we consider a model in which the radio emission originates in
a recombining wind, and the infrared component in a gaseous disk.  The
wind component has the same characteristics as in Model~B.  Matching
the mid-infrared part of the SED with a gaseous disk requires an
extinction of 31.5 mag, similar to Model~A.  The mid-infrared spectral
index is close to $\alpha = 2$, i.e.  to the value expected for an
optically thick isothermal medium The dereddened mid-IR observations
can be fitted by a blackbody of at least 5\,000 K, the flux level
constraining the size (i.e. surface area) of this component. If the
temperature is higher, the same extinction allows the near-infrared
part of the SED to be partially optically thin.

Lacking a constraint on the density gradient, we use $m = 2.0$.  We
retain the disk geometry assuming an opening angle of 5 degrees and a
temperature of 10\,000 K.  We obtain $n_{\circ} = 1.8\,\times 10^{14}$
cm$^{-3}$ and an outer radius of $R_{\rm disk} = 140.0$ \Rsun.  This
component is plotted in the bottom panel of Fig.~\ref{fig_model} with
a dashed line.  It corresponds to $M_{\rm \,disk} = 2.5\,\times
10^{-7}$ \Msun\ and a $\log(L_{\rm \,disk}/\Lsun) = 4.5$.

The infrared recombination lines are easily explained by this
model. The broad optically thick portions of the lines would originate
in the disk and the optically thin contributions would be produced in
the outer wind.  The CO band-heads likely originate from the
equatorial plane and outer part of the disk, as the high densities and
low temperatures (due to the shielding) can provide the required
physical conditions for the existence of CO and for excitation of the
band-heads (see Sect.~\ref{sec_modelA}).  This location is also in
agreement with results of CO band-head modeling by Chandler et
al. (1995) (see Sect.~\ref{sec_tool}).

In conclusion, model\,C fulfills all the observational diagnostics and
fits the observations best. Similarly to model\,A, model\,C implies a
disk luminosity that is comparable to the stellar luminosity of a B0~V
star for which $\log(L_{\rm \,disk}/\Lsun) = 4.86$ (Smith et
al. 2002).  In principle this is possible, though it implies that the
total opening angle of the disk should be $\sim$ 45 degrees. 
If we have underestimated the stellar luminosity then the disk could 
have a smaller opening angle.  Alternatively, the disk could have an
intrinsic energy source, e.g. as a result of accretion
(cf. Sect.~\ref{sec_modelA}) providing (part of) the disk luminosity.

\subsection{The nature of NGC2024/IRS2}

The location of IRS2 in the star-forming complex Orion B suggests it
is a young object, perhaps even in a pre-main sequence stage of
evolution. Objects similar to the Becklin-Neugebauer (BN) object in
Orion, which reside in cores of molecular clouds, have been proposed
to represent deeply embedded massive YSOs.  Therefore, we compare the
properties of IRS2 with those of BN-like objects (see Henning \&
G\"{u}rtler 1986 for a review).

Both IRS2 and the BN-like objects are very compact sources in the
near- and mid-IR as well as in the radio, with sizes of about 10 to
100 AU. For an \AV\ of 31.5 mag and the adopted distance of 363 pc, we
find a minimum luminosity for IRS2 of 10$^{4}$ \Lsun, corresponding to
a B0.5\,V star. This is in reasonable agreement with that of the
BN-like objects, which have integrated luminosities in the range of
$\sim$ 10$^{3}$ to 10$^{4}$ \Lsun. The near-IR hydrogen recombination
lines in BN-like objects, like in IRS2, have extended wings (with FWHM
of several hundreds of kilometers or more) and have flux ratios which
are not consistent with Case B recombination. Also, both the BN-like
objects and IRS2 show CO band-heads, implying dense and warm neutral
gas. However, the BN-like objects show evidence for an optically thick
dust cocoon as well as for a cold molecular surrounding. Both are
absent in IRS2. Perhaps IRS2 is in a similar evolutionary stage
as the BN-like objects, but somehow has its circumstellar cloud disrupted
by another more evolved source (such as IRS2b). More likely, the
star is more evolved. This would be supported by the 
apparent absence of dust in the system. If the disk in IRS2 would 
represent a remnant of the stellar formation process, one 
would expect dust to be present as it is a constituent of the
molecular cloud material that is accreted.



\section{Conclusion}
\label{sec_conclusion}

We have investigated the near-infrared and radio source
NGC\,2024-IRS2/G206.543-16.347 focusing on K-, L' and L-band
spectroscopy and on a reconstruction of the spectral energy
distribution from near-IR to the radio wavelengths. The primary goal
was to obtain better insight into the nature of the central star and
its circumstellar medium.  Simple models were used to test a number of
diagnostics, notably the slope and flux level at radio wavelengths,
the flux ratios of near-IR hydrogen lines, and the CO band-head
emission.

Previously, IRS2 has been associated with the IRAS source 05393-0156,
and, on the basis of this was found to satisfy the color-color
criteria for
\uchii\ regions. We show from beam-size arguments that the large infrared
flux peaking around 100 \mum\ does not originate from a dusty cocoon
confining the ionized gas around IRS2, but rather from a
background. This is consistent with the offset of 72.8" between IRS2
and the IRAS source 05393-0156. Therefore, IRS2 may not be interpreted
as a `cocoon' star, i.e. as the object usually depicted as an \uchii\
region.

To gain a better understanding of the nature of the circumstellar
medium we compared our observations with a number of relatively simple
models: a dense gaseous disk; a stellar wind plus a dust shell; and a
stellar wind plus a gaseous disk. The last model fits the
observational constraints best.  New spectra and the models allow one
to draw a number of conclusions concerning the properties of the
central star and of the ambient medium of IRS2.  These conclusions
are:

\begin{enumerate}

\item Both the radio flux and the 2.08-2.18 \mum\ spectrum are
      consistent with an early-B spectral type for the central star.

\item Based on a comparative study of the ice-band optical depth in
      IRS2 and in IRS2b the visual extinction towards IRS2 is
      constrained to be between 27 and 36.5 mag, with a preference for an
      \AV\ of 31.5 mag based on the shape of the SED.

\item The presence of dust cannot be completely excluded, although
      emission from dust cannot be the dominant source of radiation in
      the infrared. A Significant amount of dust can only be present in
      the equatorial plane of an optically thick gaseous disk, where
      it would be unable to contribute much to the infrared spectrum.

\item The infrared recombination lines observed in IRS2 originate for
      the most part in a high velocity (400 \kmsec) optically thick
      medium, likely distributed in a disk, and to some extent in a
      lower velocity ($\le$ 130 \kmsec) optically thin medium.

\item \mgii\ and possibly \feii\ lines are observed, implying the
      presence of a dense and warm medium with velocities of about
      (400 \kmsec), similar to the hydrogen lines.

\item The CO band-head emission originates in a high density region,
      where the temperature must drop to about 4\,000 K and where
      velocities are about 200 \kmsec. For such relatively low
      temperatures to arise close to the star requires shielding. We have
      shown that this shielding may be provided by high density gas
      located within some ten stellar radii from the star.

\item The radio data are best described by a stellar wind model with a
      mass-loss rate of $1.1\,\times 10^{-6}$ \msunyr\ recombining 
      at 145 AU. The radio emission is variable 
      on short timescales (months), its spectral index might also vary.
      These variations may be explained by changes in the ionization
      and/or density structure; however, non-thermal radiation 
      cannot be excluded.

\item The origin of the infrared SED is best described by a high 
      density disk-like gaseous medium that is optically thick 
      in the mid-infrared and extends to about 15 \rstar.
            
\item 
      The luminosity emitted by the circumstellar material is
      comparable to the total luminosity of an early-B main-sequence
      star.  It is possible that either the star has a higher
      luminosity, which would be consistent with the high mass-loss
      rate derived from radio observations, or part of the luminosity
      is supplied by the disk e.g. through accretion.  

\end{enumerate}

\begin{acknowledgements}
   We would like to thank the Observatory Staff of UKIRT and of the
   VLT.  We are grateful to L. Decin for her help in the data
   reduction, to S. Hony and C. Neiner for useful and repeated
   discussions.  M. Marlborough and A. Sigut are acknowledged for
   their help with the temperature model. We thank R. Gaume for his willingness
   to share the archival VLA data. UKIRT is operated by the 
   Joint Astronomy Centre on behalf of the U.K. Particle Physics and
   Astronomy Research Council. TRG's research is supported by the Gemini
   Observatory, which is operated by the Association of Universities for
   Research in Astronomy on behalf of the international Gemini partnership
   of Argentina, Australia, Brazil, Canada, Chile, the United Kingdom, and
   the United States of America.

\end{acknowledgements}
                        
{}

\appendix
\section{}
\begin{table*}[t!]
\begin{center}
\caption[]{Small aperture data used to construct the spectral energy 
           distribution: we list the observing date, frequency,wavelength, beam 
           size, magnitude, flux, observational technique when not 
           photometric, and reference.
           \label{tab_short}}
\begin{tabular}{rccccccc}
Date   & $\nu$&$\lambda$&Beam &Magnitude&Flux&Tech & Ref  \\
       & $10^{12}$ Hz &(\mum)    &(\arcsec)&    &(Jy)    &     & \\ 
\hline  
1976-77 & 326   &  0.92     &                  &$\geq$ 16.5    &   $\leq$0.001        &(a)& F79 \\     
1974    & 240   &  1.25     & $\leq$12         &10.80$\pm$0.10 &   0.072$\pm$   0.007 &   & G74 \\     
1982    & 240   &  1.25     &       10         &11.64$\pm$0.05 &   0.033$\pm$   0.002 &   & B89 \\     
1986    & 240   &  1.25     &       15         &11.57$\pm$0.02 &   0.036$\pm$   0.001 &   & C86 \\     
1991    & 240   &  1.25     &       15         &10.56$\pm$0.05 &   0.090$\pm$   0.004 &   & N94 \\     
1999    & 240   &  1.25     & 0.7$\times$0.7 &11.02$\pm$0.15 &   0.059$\pm$   0.008 &(b)& B03 \\
1974    & 182   &  1.65     & $\leq$12         & 7.32$\pm$0.05 &   1.226$\pm$   0.056 &   & G74 \\     
1982    & 182   &  1.65     &       10         & 7.38$\pm$0.05 &   1.160$\pm$   0.053 &   & B89 \\     
1986    & 182   &  1.65     &       15         & 7.38$\pm$0.02 &   1.160$\pm$   0.021 &   & C86 \\     
1991    & 182   &  1.65     &       15         & 6.53$\pm$0.05 &   2.538$\pm$   0.117 &   & N94 \\   
1974    & 136   &  2.20     & $\leq$12         & 4.54$\pm$0.05 &  10.423$\pm$   0.480 &   & G74 \\     
1982    & 136   &  2.20     &       10         & 4.59$\pm$0.05 &   9.954$\pm$   0.459 &   & B89 \\     
1986    & 136   &  2.20     &       15         & 4.57$\pm$0.02 &  10.139$\pm$   0.187 &   & C86 \\     
1991    & 136   &  2.20     &       15         & 3.62$\pm$0.05 &  24.322$\pm$   1.120 &   & N94 \\    
1993    & 136   &  2.20     & 3.1$\times$3.1   & 4.7 $\pm$0.2  &   8.995$\pm$   1.666 &   & N94 \\     
1994    & 136   &  2.20     & 3.1$\times$3.1   & 4.5 $\pm$0.2  &  10.814$\pm$   2.003 &   & N94 \\     
1999    & 136   &  2.20     & 0.7$\times$0.7 & 4.75$\pm$0.1  &   8.590$\pm$   0.792 &(b)& B03 \\
1997    & 86-115   &  2.6-3.5  & 14$\times$20     &               &   25-35              &(c)& ISO \\ 
1974    & 100   &  3.00     & $\leq$12         & 3.76$\pm$0.05 &  13.115$\pm$   0.604 &   & G74 \\     
1974    & 98.3   &  3.05     & $\leq$12         & 4.08$\pm$0.05 &   9.506$\pm$   0.438 &   & G74 \\     
1974    & 88.2   &  3.40     & $\leq$12         & 2.54$\pm$0.05 &  32.779$\pm$   1.510 &   & G74 \\     
1982    & 79.7   &  3.76     &       10         & 2.06$\pm$0.05 &  42.973$\pm$   1.980 &   & B89 \\     
1986    & 79.7   &  3.76     &       15         & 1.95$\pm$0.04 &  47.554$\pm$   1.752 &   & C86 \\     
1993    & 79.7   &  3.76     & 3.1$\times$3.1   & 1.7 $\pm$0.2  &  59.867$\pm$  11.090 &   & N94 \\     
1994    & 79.7   &  3.76     & 3.1$\times$3.1   & 1.9 $\pm$0.2  &  49.796$\pm$   9.225 &   & N94 \\
1991    & 78.3   &  3.83     &       15         & 0.62$\pm$0.05 & 156.809$\pm$   7.224 &   & N94 \\     
1997    & 37.5-74.9   &  4.0-8.0  & 14$\times$20     &               &   40-70              &(c)& ISO \\   
1982    & 64.6   &  4.64     & 0.27$\times$5    & 1.09$\pm$0.03 &  72.489$\pm$   2.003 &(d)& J84 \\  
1986    & 62.5   &  4.80     &       15         & 1.20$\pm$0.04 &  61.609$\pm$   2.270 &   & C86 \\ 
1974    & 62.5   &  4.80     &   $\leq$12       & 1.24$\pm$0.05 &  59.380$\pm$   2.736 &   & G74 \\ 
1974    & 35.7   &  8.40     &   $\leq$12       & 0.80$\pm$0.05 &  30.367$\pm$   1.399 &   & G74 \\ 
1974    & 29.4   & 10.2      &   $\leq$12       & 1.62$\pm$0.05 &   9.552$\pm$   0.440 &   & G74 \\ 
1997    & 25.0-29.4   & 10.2-12.0 & 14$\times$20     &               &   15-35              &(c)& ISO \\ 
1998    & 28.6   & 10.5      &       1.5        &               &  19.000$\pm$   1.000 &   & W01 \\ 
1974    & 26.8   & 11.2      &   $\leq$12       & 1.06$\pm$0.05 &  13.119$\pm$   0.604 &   & G74 \\ 
1997    & 15.4-25.0   & 12-19.5   & 14$\times$27     &               &   35-80              &(c)& ISO \\ 
1974    & 2.3.8   & 12.6      &   $\leq$12       & 0.54$\pm$0.05 &  16.416$\pm$   0.756 &   & G74 \\ 
1974    & 1.5.0   & 20.0      &   $\leq$12       & 0.19$\pm$0.04 &   7.925$\pm$   0.292 &   & G74 \\ 
1974    & 1.3.3   & 22.5      &   $\leq$12       & 0.09$\pm$0.06 &   6.562$\pm$   0.363 &   & G74 \\ 
1986    & 0.231   & 1300      & 11               &               &       $\le$0.2       &(e)& M88 \\
1994    & 0.111   & 2700      &8.2$\times$7.7    &               &0.1600 $\pm$0.0140    &(e)& W95 \\ 
\end{tabular}
\begin{list}{}{}
\item[Tech:] 
(a): Photography;
(b): Narrow-band photometry;
(c): Spectroscopy;
(d): Speckle interferometry;
(e): Interferometry. 
 \item[Ref:] 
G74: Grasdalen 1974; 
J84: Jiang, Perrier \& L\'ena 1984;
C86: Chalabaev \& L\'ena 1986; 
S86: Snell \& Bally 1986;  
B89: Barnes et al. 1989; 
G92: Gaume, Johnston \& Wilson 1992;
K94: Kurtz, Churchwell \& Wood 1994; 
N94: Nisini et al. 1994;
W95: Wilson, Mehringer \& Dickel 1995; 
W98: Walsh et al. 1998; 
W01: Walsh et al. 2001;
B03: Bik et al. 2003;
ISO: {\em Infrared Space Observatory}, obtained in AOT6 mode.
\end{list}
\end{center}
\end{table*}

\begin{table*}[t]
\begin{center}
\caption[]{Small aperture interferometric data used to construct the 
           spectral energy distribution at centimeter wavelengths. The
           listed quantities are the same as in Tab.~\ref{tab_short}.
           \label{tab_short_cm}}
\begin{tabular}{rlllcc}
    Date   & $\nu$&$\lambda$&Beam      &  Flux            & Ref \\
           &  GHz &(cm)   & (\arcsec)&  (mJy)           &     \\ 
\hline  
  Aug 1994 &23   & 1.3  &  0.3         &  88.8 $\pm$ 8.9  & VLA \\
  Oct 1994 &23   & 1.3  &  0.9         &  74   $\pm$ 7    & VLA \\          
  Jan 1995 &23   & 1.3  &  2.8         &  65   $\pm$ 7    & VLA \\  
  Apr 1995 &23   & 1.3  &  2.8         &  58   $\pm$ 15   & VLA \\           
  Aug 1995 &23   & 1.3  &  0.08        &  95   $\pm$ 10    & VLA \\  
  Nov 1989 &23   & 1.3  &  2.4         &  14.7 $\pm$ 10.0  & G92 \\    
  Mar 1989 &15   & 2    &  0.5         &  19.5 $\pm$ 0.4  & K94 \\           
  May 1994 &15   & 2    &  0.4         &  40   $\pm$ 5    & VLA \\		 
  Aug 1994 &15   & 2    &  0.4         &  47.2 $\pm$ 4.7  & VLA \\
  Oct 1994 &15   & 2    &  1.2         &  48   $\pm$ 5    & VLA \\
  Jan 1995 &15   & 2    &  3.9         &  48   $\pm$ 7    & VLA \\  
  Apr 1995 &15   & 2    &  3.9         &  42   $\pm$ 5    & VLA \\
  Aug 1995 &15   & 2    &  0.14        &  40   $\pm$ 4.0  & VLA \\  
  Mar 1989 &8.4  & 3.6  &  0.5         &   8.9 $\pm$ 0.2  & K94 \\     
      1994 &8.4  & 3.6  &  1.0         &   6.0 $\pm$ 1.0  & W98 \\     
  May 1994 &8.4  & 3.6  &  0.7         &  26   $\pm$ 4    & VLA \\   		 
  Aug 1994 &8.4  & 3.6  &  0.7         &  27.5 $\pm$ 2.8  & VLA \\   
  Oct 1994 &8.4  & 3.6  &  2.3         &  22   $\pm$ 10   & VLA \\  
  Jan 1995 &8.4  & 3.6  &  8.4         &  20   $\pm$ 10   & VLA \\
  Aug 1995 &8.4  & 3.6  &  0.24        &  28.1 $\pm$ 2.8  & VLA \\    
  Mar 2002 &8.4  & 3.6  &  0.24        &  17.4 $\pm$ 0.6  & R02 \\ 
  Nov 1983 &5    & 6    &  2.6         &   9.7 $\pm$ 2    & S86 \\           
  May 1994 &5    & 6    &  1.2         &  13   $\pm$ 2    & VLA \\    		 
  Aug 1994 &5    & 6    &  1.2         &  16   $\pm$ 1.6  & VLA \\
  Aug 1995 &5    & 6    &  0.4         &  14   $\pm$ 1.4  & VLA \\     
\end{tabular}
 \begin{list}{}{}
 \item[Ref:] 
  S86: Snell \& Bally 1986;
  G92: Gaume, Johnston \& Wilson 1992;
  K94: Kurtz, Churchwell \& Wood 1994; 
  R02: Rodr\'\i guez (priv.  comm. 2002);
  W98: Walsh et al. 1998;
  VLA: {\em Very Large Array} Archival Data
 \end{list}
\end{center}
\end{table*} 

\begin{table*}[h]
\begin{center}
\caption[]{Large aperture data used to construct the spectral energy distribution: 
           we list the same quantities as in Tables~\ref{tab_short} and~\ref{tab_short_cm}.
           \label{tab_long}
          }
\begin{tabular}{rcccccc}
Date     & $\nu$ & $\lambda$ &   Beam          &      Flux        & Tech  & Ref  \\
         & 10$^{12}$  Hz  &  (\mum)   &  (\arcsec)      &  (Jy)            &       &      \\ 
\hline
1976-77  &136&  2.20     &  62             & 144.76$\pm$  3.55&       & F79  \\  
1976-77  &85.7&  3.50     &  62             & 302.3 $\pm$  3.55&       & F79  \\    
1983     &25.0&  12       &  45$\times$270  &  284.4$\pm$ 17.0 &       & IRAS \\           
1983     &12.0&  25       &  45$\times$276  &   4746$\pm$570   &       & IRAS \\    
1997     &10.3-10.5&  28.6-29  &  20$\times$27   & 325-360          & (c)   & ISO  \\     
1997     &7.89-10.3&  29-38    &  20$\times$33   & 520-940          & (c)   & ISO  \\            
1981-83  &7.50&  40       &  49             &   1640$\pm$410   &       & T84  \\            
1997     &1.52-6.97&  43-197   &  84             &   5000-15000     & (c)   & ISO  \\   
1981-83  &5.00&  60       &  49             &   3600$\pm$1200  &       & T84  \\            
1983     &5.00&  60       &  90$\times$282  &   7894$\pm$1200  &       & IRAS \\           
1981-83  &3.00&  100      &  49             &   3600$\pm$1200  &       & T84  \\            
1983     &3.00&  100      &  180$\times$300 &       $\le$35330 &       & IRAS \\           
1981-83  &1.87&  160      &  49             &   1710$\pm$570   &       & T84  \\            
1985     &0.857&  350      &  90             &     95$\pm$20    &       & M88  \\            
1985     &0.231&  1300     &  90             &     22$\pm$2     &       & M88  \\    
\end{tabular}
 \begin{list}{}{}
\item[(1)] 
(c): Spectroscopy
 \item[(2)] 
T84: Thronson et al. 1984; 
F79: Frey et al. 1979;
M88: Mezger et al. 1988;
IRAS: {\em Infrared Astronomical Satellite};
ISO: {\em Infrared Space Observatory}.
\end{list}
\end{center}
\end{table*}

\end{document}